# Observation of electric current induced by optically injected spin current

Xiao-Dong Cui[1,2*], Shun-Qing Shen[1,3*], Jian Li[1], Weikun Ge[4], Fu-Chun Zhang[1,3]

[1]*Department of Physics, The University of Hong Kong, Hong Kong, China*

[2]*Department of Chemistry and HKU-CAS Joint Laboratory on New Materials, The University of Hong Kong, Hong Kong, China*

[3]*Center of Theoretical and Computational Physics, The University of Hong Kong, Hong Kong, China*

[4]*Department of Physics and Institute of Nano-Science and Technology, The Hong Kong University of Science and Technology, Clear Water Bay, Hong Kong, China*

*To whom correspondence should be addressed. E-mail: X. D. Cui. (xdcui@hkucc.hku.hk) or S. Q. Shen (sshen@hkucc.hku.hk).

*Abstract*

**A normally incident light of linear polarization injects a pure spin current in a strip of 2-dimensional electron gas with spin-orbit coupling. We report observation of an electric current with a butterfly-like pattern induced by such a light shed on the vicinity of a crossbar shaped InGaAs/InAlAs quantum well. Its light polarization dependence is the same as that of the spin current. We attribute the observed electric current to be converted from the optically injected spin current caused by scatterings near the crossing. Our observation provides a realistic technique to**

**detect spin currents, and opens a new route to study the spin-related science and engineering in semiconductors.**

Coherent transport and control of electron spins in semiconductor hetero-junctions have been studied intensively with the ultimate aim to implement spintronic devices. (1-3) Electric control of quantum spin in semiconductors may be realized through a relativistic effect which couples an electron's orbital motion to its spin orientation. The spin-orbit coupling can be enhanced drastically in semiconductors due to the band structures of electrons. A two-dimensional electron gas (2DEG) in heterostructure with $C_{2v}$ symmetry may have a vertical structural inversion asymmetry (SIA), which induces a Rashba spin-orbit coupling and leads to spin splitting of the conduction band in the momentum k-space. (4,5) This system provides a good platform to control electron spin in semiconductors by electric fields. Examples include spin photocurrent, (6) electric-dipole-induced spin resonance, (7) and spin coherent transport. (8-12) In this system, conduction electrons optically pumped from the valence band will generate a spin polarized electric current, whose direction and magnitude depend on the helicity and the incident angle of light as well as the band structure of electrons. This spin photo-galvanic effect has been studied extensively and observed experimentally. (13,14) In the case that the incident light is normal to the two-dimensional plane, electrons with opposite spins travel in opposite directions. As a result, a pure spin current circulates while the electric current vanishes. Detecting spin current, however, has been a challenging problem. (17-20) Here we report an experimental observation of electric currents in a crossbar shaped two-dimensional InGaAs/InAlAs system induced by the pure spin current injected with linearly polarized lights. The observed transverse currents have a butterfly-like pattern: either both inward or both outward, as shown in Fig. 1a. This is distinguished from the conventional Hall current. The light polarization dependence of the electric current is in an excellent agreement with the expected dependence of the spin current. Our observations provide a realistic technique to detect spin currents, and open a new route to study the spin-related science and engineering in semiconductors.

Let us start with a brief introduction of the pure spin current induced by a light normally incident on the x-y plane of a $C_{2v}$ 2EDG with a Rashba coupling $H_{SIA}=-\lambda(k_x\sigma_y - k_y\sigma_x)$, with k the momentum of electron and $\lambda$ the spin-orbit coupling constant. (21) A pure spin current is defined by $j_{\alpha\beta}= (v_\alpha\sigma_\beta +\sigma_\beta v_\beta)/2$, with v the velocity and $\sigma$ the spin Pauli matrices of electron. The states with equal energy in the conduction band are represented by a pair of spin-dependent concentric circles of radii $k_+$ and $k_-$. For example, for an electron with k along the x-axis, the four degenerated states are $|k_+,\uparrow>$, $|k_-,\downarrow>$, $|-k_-,\uparrow>$, and $|-k_+,\downarrow>$, respectively, where $|\uparrow>$ and $|\downarrow>$ are the two eigenstates of spin $\sigma_y$, respectively, and $k_+ - k_- =2m^*\alpha/\hbar^2$ with m* the effective mass. The two degenerate states on the same concentric circle, for example $|k_+,\uparrow>$ and $|-k_+,\downarrow>$, have opposite velocities $v_\pm=\pm(\hbar k_+/m^*-\alpha/\hbar)$, but carry equal spin current, $j_{xy}= (\hbar^2 k_+/m^*-\alpha)/2$. Thus the pair contributes a null electric current, but a finite spin current of $\hbar^2 k_+/m^*-\alpha$. The total spin current is determined by the band structure and optical excitation processes as illustrated in Fig.1b. In a steady optical excitation process, the spin current is proportional to the spin relaxation time and the transition rates to the bottom of spin-dependent conduction bands. (22) The contribution from the holes in valence bands is negligible since their spin relaxation time is typically short. (23) Assuming the inter-band depahsing is fast, the injection rate of spin current can be calculated using the Fermi Golden rule (16) or the solution of semiconductor optical Bloch equations (24). A linearly polarized light can be decomposed as a combination of two circularly polarized beams of light. Assume the angle $\Phi$ between the polarization plane and the y-axis, the phase difference between these two composite beams of the light is 2Φ. A detailed calculation (16) for a linearly polarized light gives the spin current $J_{xy}=J_0+J_1\cos2\Phi$. Either left or right circularly polarized light may pump electrons from the valence to conduction band, and generate a spin current. The polarization dependence of the spin current originates from the interference of two composite circularly polarized lights.

Physical picture of spin current generated in this process is quite simple and straightforward. The crux is on observable physical effects produced by the spin current. This is the motivation of the present work. For this aim we designed an experimental setup, as shown in Fig. 2a, which is carved on a sample of 2DEG formed at $In_{0.65}Ga_{0.35}As/In_{0.52}Al_{0.48}As$ interfaces of the quantum well with thickness of 14nm, grown on thick semi-insulating (001) InP substrates by Molecular Beam Epitaxy as described elsewhere. (25) The SIA was achieved by δ–doping on the top of the well. The carrier density and mobility are about $1.5\times10^{12}cm^{-2}$ and $1.1 \times 10^5$ $cm^2/Vs$ determined by the Hall measurement, and the Rashba coupling is about $3.0\times10^{-12}eVm$ by the Shubnikov-de Hass oscillation. The sample has two electric channels of widths 20 μm and 200 μm respectively, with one narrower and one wider than the assumed spin coherence length to check the size asymmetric effect of the optically injected spin current. We also carried out measurements on similar structured InGaAs/InAlAs 2DEGs with a graded In composition from x = 0.53 to 0.75 in the QW, with a Rashiba coupling about 6.3x10-12eVm. Essentially same results are observed. The planar electric channels were carved out by the standard photolithography and wet etching along [110] or [1$\bar{1}$0] defined as lab x- and y-axis. (see Figure 2a). Electric contact was made to 2DEG with Ni/Au metal electrodes at far ends of electric channels. A linearly polarized laser light of wavelength 880 nm and power of 40 mW from a tuneable Ti:sapphire laser was selected as the excitation source for the inter-band transition. The incident light was first fed through a photo-elastic modulator which modulates its polarization oscillating between the two orthogonal directions with a period of 50 KHz. Then the light was focused through a 10× objective lens to normally shed on a 10 μm-diameter spot on the sample which was mounted in a cryostat at liquid nitrogen temperature. Electric currents passing through all terminals were monitored by voltage drops at symmetrically loaded resistors of 633Ω. The voltages were read out with a lock-in amplifier which is locked at 50 KHz so that the electric currents induced by non-polarization related phenomena e.g. Dember effect and thermo-electric effects were eliminated clearly.

No measurable current along the wide channels (y-axis) was observed on the background noise of tens of pico ampere under normal incidence, though a significant photocurrent was observed if the circular polarized light was shed at an oblique angle. (25) However, when a light with linear polarization along the x- or y-axis was normally incident to the plane near the electric channel junction or the adjacent, we observed reproducible currents along the narrow channels in the unbiased conditions. The pattern of the current flow near the junction of A-C and B-C is plotted in Fig. 2b. Figs. 2c-e show the currents passing through the terminals as functions of the spot position of the incident light along the 20μm-wide channel (x-axis). The current peaks at x = 0 μm and 200 μm corresponding to the positions of the two channel junctions, and vanishes at a distance of 50 μm away from the peak position. Note that the current curves of $I_{C-A}$ and $I_{C-B}$, and $I_{A-D}$ and $I_{B-D}$ almost overlap. Namely, the current flows inward or outward simultaneously through the channel junction. The spatial distribution of the observed current is substantially different from the current in the Hall effect and reciprocal spin Hall effect, where the Hall current is unidirectional. *The inward and outward butterfly-like pattern of the currents and the absence of an electric current along the wide channel are the* *signature of the observed effect as summarized in Fig. 1a.* The disappearance of all currents at the middle point of the channel C and D, x = 100 μm, indicates the observed phenomena are not conventional photocurrents. Note that the spatial location of the light spot indicated in Figure 2 is hundreds of microns away from the metal electrodes. Therefore the contribution to the measured electric current from the interface between metal electrodes and semiconductor channels is negligible.

As the current measurements were locked at the polarization oscillating frequency, the obtained current measures the difference between the electric currents generated by the two orthogonally polarized lights in this device. To further examine the relation between the observed currents and polarization of the incident light, we rotate the polarization of the incident lights simultaneously while the light spot is fixed at the peak position of $I_{C-A}$ and $I_{C-B}$ in Fig. 2. The current varies with the angle Φ between the light polarization and the y-axis, and a good behaviour of cos2Φ is shown in Fig. 3. The light polarization dependence of the current is excellent agreement with the angle dependence of the spin current predicted theoretically. Note that the lock-in technique we used does not measure any Φ independent part of the current because of the cancellation.

The result of the laser scanning along the edge of the channel A-B at x = 10 μm is shown in Fig. 4. Electric currents $I_{A-C}$ and $I_{B-C}$ were observed near the channel junction, and they were almost identical. The spatial dependence of the current are drastical in the vicinity of Y=0, or the center of the channel CD. The current $I_{A-B}$ remains negligible at all the laser spot position. No current in the channel D is detected as the distance between the channel and the light spot is too far, well beyond the spin mean free path.. When the laser scans along the mid line of the channel A-B, *i.e.*, at x = 100 μm, no current was observed, for the light spot is located well beyond the spin coherence distance. This is consistent with the conventional spin galvanic effect, where the current vanishes when the linearly polarized light sheds normally on the plane. We conclude that the observed electric current near the junction with a narrow width (20 μm) is not a conventional photocurrent. (13, 14, 25) The equality of $I_{A-C}$ and $I_{B-C}$ as shown in Fig. 4 obviously excludes the possibility of the diffusion mechanism. It indicates that the observed current is a scattering effect near the junction as the spin coherence length is expected to be order of 20 μm in the present experiment.

We interpret the observed electric current as a result of the the scattering of the spin current near the junction. In this view, the optical excitation induced by the incident light is considered to be a source of the spin current, whose spin polarization aligns within the plane and is perpendicular to the direction of motion of the spin. The spin current is given by $J_{xy} = J_0 + J_1 \cos 2\Phi$. The same dependences on the light polarization of the electric current and $J_{xy}$ shown in Fig. 3 can be considered as a direct evidence in strong support of this scenario. Furthermore, the sample is carved on a uniform strip of InGaAs/InAlAs, so that the spin injection problem is clearly avoided when the spin current passes through the scattering regime at the junction. A typical value of electron spin decoherence time is several hundreds of ps (26) and the Fermi velocity of the sample is $v_F \sim 4.4 \times 10^4$ m/s at the carrier density $1.5 \times 10^{12}$ cm$^{-2}$. Thus the spin coherence length is roughly estimated to be tens of μm, which, as a result, is qualitatively consistent with the width of the peaks in Fig. 2c-e and Fig.4. Here the spin coherence length at liquid nitrogen temperature is longer than the widths of 20 μm of the two narrow channels. The impurity effect may not be crucial in this experiment, but the phase coherence length of electron is much shorter than 20 μm.

The conversion of spin current to electric current has been discussed extensively in the context of the reciprocal spin Hall effect, where the spin current is polarized along the direction perpendicular to the plane and the effect is attributed to the Mott asymmetric scattering of electrons. (27) In the present case the spin current is polarized within the plane, and perpendicular to the direction of motion of the spin. As examined in Ref. 16, an electron of spin-y polarization will be equally scattered along the x and −x axes. As a result the scattered electric current has a butterfly-like pattern: either both outward or both inward, in the transverse directions of the spin current as shown in Fig. 1a. Our experimental observation of the electric currents is in excellent qualitative agreement with this picture. In the Landauer-Buttiker formalism, (28) numerical simulations were performed on a symmetric crossbar shaped setup in the tight binding approximation. The numerical results and symmetry analysis show that the resulting electric current depends strongly on the spin polarization of the spin current. (29) For the case of the in-plane polarized spin current as in the present experiment, the conversion rate for the spin current to electric current is estimated to be 0.3% ~ 1%. (16) In the present work, though there are some unknown parameters of the sample, the amplitude of the injected spin current $J_{xy}$ by the linearly polarized light can be roughly estimated by using the measurable photocurrent $J_x$ injected by the circularly polarized light at a small oblique angle $\Theta$ away from the normal position. If the powers of the applied two laser lights are equal, a theoretical calculation gives $J_x(\Theta)/J_{xy} \approx 0.04\ \Theta\ (2e/\hbar)$ by using the solution of semiconductor Bloch optical equation near the $\Gamma$ point of 2DEG of InGaAs. Taking into account the size of light spot, geometry of the setup and laser power, we estimates $J_x(\Theta)$ ~ 94 $\Theta$ nA, from the experimental data.(25), hence the injected spin current is estimated to be approximately $2.3 \times 10^3$ nA ($\hbar/2e$). The observed electric current in the present experiment is several of nA, say 4.0nA for the peaks in Fig. 2c. The conversion rate from

9spin current to charge current is roughly about 0.16 %, which is close to the numerically simulated values. Besides we wish to emphasize that the injection procedure of the spin current breaks the time reversal symmetry, and the observed current should be dissipative.

In summary, we have observed an electric current of butterfly-like pattern near the junction of a cross-bar shaped (001) InGaAs/InAlAs quantum well sample with a linearly polarized light incident perpendicularly to the sample surface. The pattern and the light polarization dependence of the electric current are in agreement with the scenario that the electric current is induced by scattering of the optically injected spin current near the cross-bar. The observed phenomenon is different from the reciprocal spin Hall effect and conventional spin galvanic effect, and may provide a new route to measure and control quantum electrons in semiconductors.

22. As the energy of photons in this experiment is higher than the energy gap, the excited electrons will first decay to the bottom of spin-dependent conduction bands, and then recombine with holes in valence band. The populated spin-

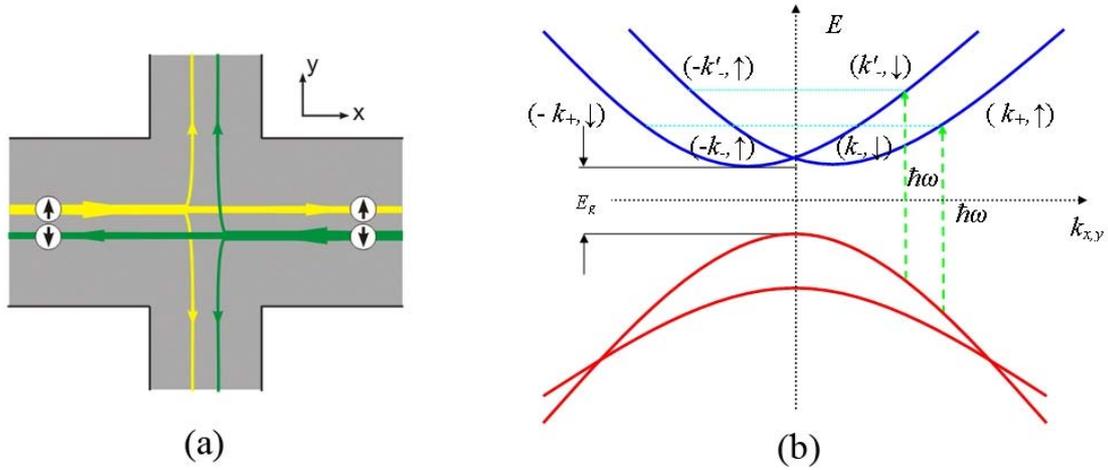

(a)  (b)

**Figure 1 (a).** The main experimental observations of the present work for a symmetric setup are summarized schematically in Figure 1a. Electrons with spin-up and -down moving in opposite directions circulate a *spin current*. Electric currents (yellow and green arrowed lines) can be induced when a spin current passes through a scattering regime with Rashba coupling. The transverse currents flow either inward or outward which is determined by the sign of Rashba coupling, and the conventional Hall current or the Hall voltage do not appear. In the present experiment an asymmetric setup is chosen to examine the size effect. **(b)** Schematic band structures of two-dimensional electron system (2DEG) and physical origin of optically injected spin current of electrons by linearly and circularly polarized lights. In this experiment linearly polarized lights are used to eliminate or substantially reduce the background noise for the polarization dependence of the induced spin current.

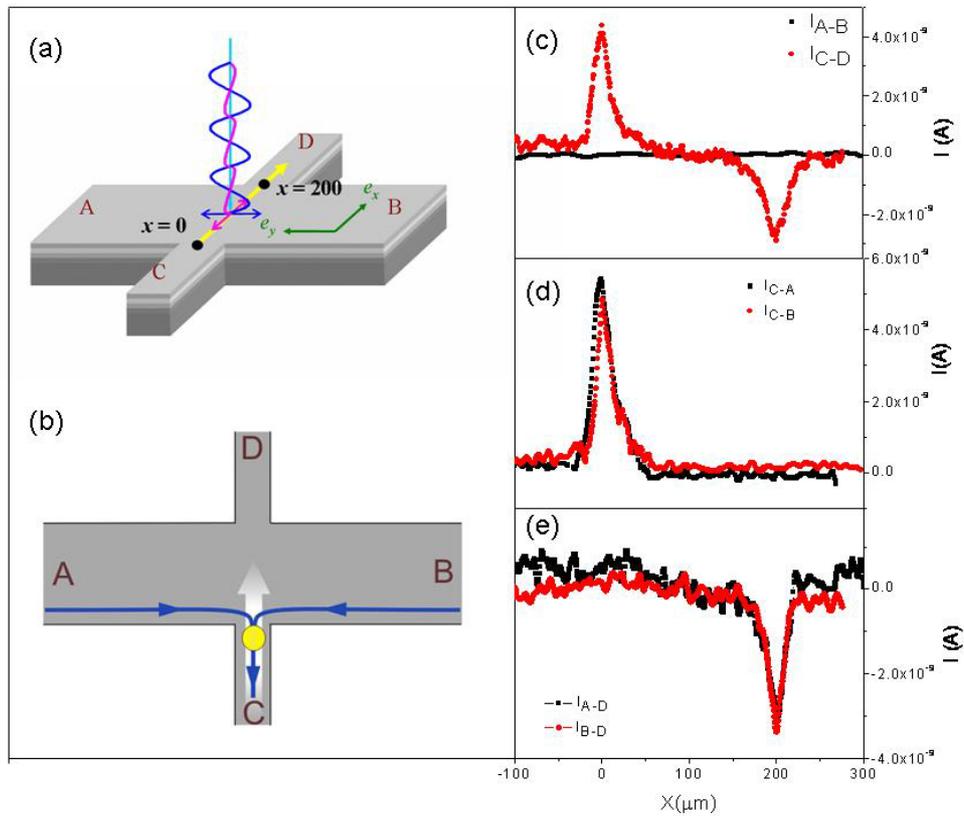

**Figure 2** (a) Schematic of the 2DEG sample and the experimental setup. The 20 μm- and 200 μm-wide planar channels are carved out by standard photolithography and wet etching along $[110]$ or $[1\bar{1}0]$ defined as lab x- and y-axis. A linearly polarized light beam, red colour for the polarization along x-axis and blue colour for the polarization along y-axis, scans perpendicularly on the 20 μm-wide channel (x-axis), where the two junctions sit at x = 0 and x = 200 μm, respectively. (b) Schematic pattern of spin currents (white arrow) generated by the light spot (yellow spot) and induced charge currents(blue lines) distribution. (c), (d) and (e) are the typical charge currents through all the terminals as a function of the light spot position at x-axis.

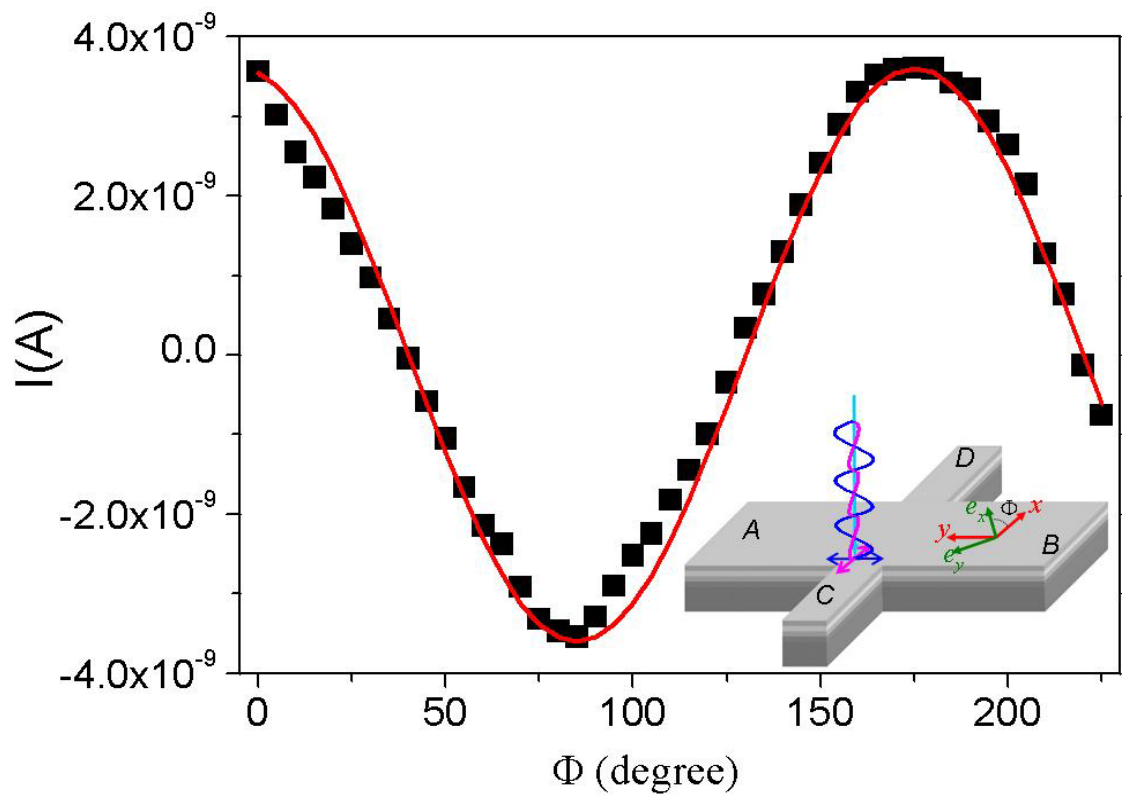

**Figure 3** The charge current through terminal A-C $I_{A-C}$ (black dot with error bars) at channel junction x = 0 varies with the polarization angel Φ relative to the lab frame, as shown in the inset. The red curve is a fit of cos2Φ.

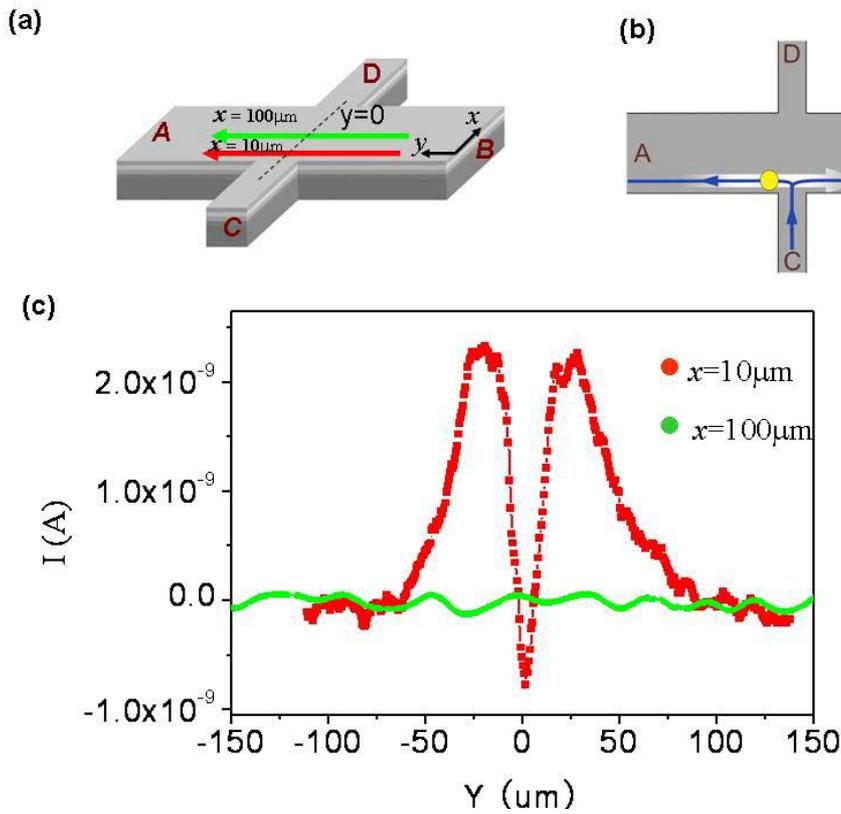

**Figure 4** (a) Schematic of the light scanning on the 200 μm-wide channel. The red and green scanning lines are parallel to y-axis and keep 10 μm and 100 μm away from the channel edge respectively. (b) Schematic pattern of spin currents (white arrow) generated by the light spot (yellow spot) and induced charge currents (blue lines) distribution. (c) The corresponding charge currents $I_{A-C}$ show strong position dependence, where the junction centres at y = 0. The currents in the channels A-C and B-C are almost equal $I_{A-C} \approx I_{B-C}$ while the current in the channel A-B are measured negligibly no matter where the light spot is. The slight asymmetry of the current about y = 0 may be caused by the non-equal distances of the light spot to the edge.